\documentstyle[aps,prl,epsfig, twocolumn]{revtex}

\begin{document}
\title{Squeezing and entanglement of atomic beams}
\author{L.-M.~Duan$^1$, A.~S\o rensen$^2$ J.~I.~Cirac$^1$ \& P.~Zoller$^1$}
\address{$^1$ Institute for Theoretical Physics, University of
Innsbruck, A-6020 Innsbruck, Austria \\
$^2$ Institute of Physics and Astronomy, University of
Aarhus, DK-8000 \AA rhus C, Denmark\\
}
\maketitle

\begin{abstract}
We propose and analyze a scheme for generating entangled atomic beams out of
a Bose-Einstein condensate using spin--exchanging collisions. In particular,
we show how to create both atomic squeezed states and entangled states of
pairs of atoms.

\end{abstract}

The recent experimental achievement of Bose-Einstein condensates \cite{1}
has raised a lot of interest since it may lead to important applications 
\cite{2,3,4}. Some of these applications are based on the analogies between
an atomic condensate and a single mode optical field. For example, atoms in
the condensate can be outcoupled to produce a coherent atomic beam \cite{5}
similar to a laser beam. In this letter we build on these analogies to show
how to produce entanglement between atoms in different internal states
similar to the one created for photons with different polarizations by
parametric down conversion \cite{NOPA,paradown}. In particular, we analyze a
physical situation which gives rise to: (i) a beam of atoms in a broad-band
two-mode squeezed state (a continuous variable entangled state) with respect
to two internal levels; (ii) a pair of outgoing atoms in an effective
maximally entangled state in a two-dimensional Hilbert space. The physical
mechanism responsible for this processes is spin exchanging collisions,
where two atoms of the condensate interact to create two correlated atoms in
two different internal states.

We consider a Bose-Einstein condensate confined in an optical trap and in
some internal level $|0\rangle $. Two atoms in the condensate can collide to
create a pair of atoms in two other internal levels 
\begin{equation}
2|0\rangle \rightarrow |+1\rangle +|-1\rangle .  \label{eq1}
\end{equation}
The situation we have in mind is illustrated in Fig.\ 1. Levels $|0,\pm
1\rangle $ could correspond to the hyperfine Zeeman levels $|F=1,M=0,\pm
1\rangle $ of an Alkali atom. In that case, the selection rules would
prevent other collisional processes to occur. We assume that the process (%
\ref{eq1}) can be switched on and off by changing some external parameter.
For example, the condensate level could be shifted in a time-dependent way
by an external field so that energy conservation effectively allows or
inhibits the process in Eq. (\ref{eq1}). This level shift could be accomplished, for example,
by an off-resonant microwave or laser field with an appropriate
polarization. We also assume a one--dimensional situation where the trapping
potential along the transverse direction is sufficiently tight so that the
atomic motion is frozen along the $y,z$ directions. We will choose different
trapping potentials along the $x$ direction in order to illustrate our ideas
of how to create squeezed atomic states and entangled pairs. In both cases,
the potential will be identical for the atomic levels $|\pm 1\rangle $ and
such that the atoms in those levels can escape from the trap, in order to
facilitate measurements with them without being affected by the atoms in the
condensate level $|0\rangle $. For instance, this could be obtained if the optical trap
is made by a strong laser along the $x$ direction crossed by a weak laser
along $y$. In this configuration, if the energy shift between the 
$|0\rangle$ and $|\pm 1\rangle$ levels is larger than the trap depth induced by the
strong laser and smaller than that by the weak laser, the $|M_F=\pm1\rangle$ atoms will be free 
to move along $x$ but will be bound in the other two directions.
\begin{figure}[tbp]
\epsfig{file=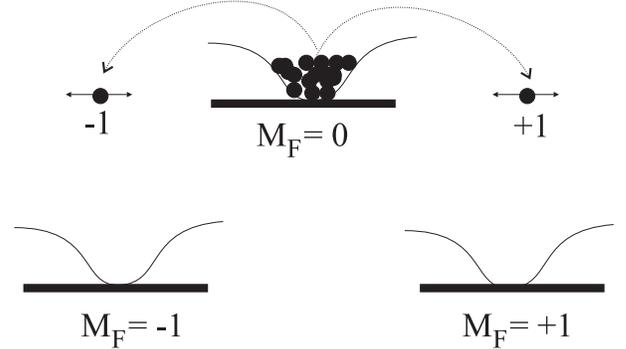,width=8cm}
\caption{Atomic configuration: The condensate is in the Hyperfine Zeeman
level $|F=1,M=0\rangle$, which is shifted with respect to the $|F=1,M=\pm
1\rangle$ states.}
\end{figure}

This situation is described by the following second quantized Hamiltonian ($%
\hbar =1$) 
\begin{eqnarray}
H &=&\sum_{i=\pm 1}\int_{-\infty }^{\infty }\hat{\phi}_{i}^{\dagger }(x)%
\left[ -\frac{\partial _{xx}^{2}}{2m}+V\left( x\right) \right] \hat{\phi}%
_{i}(x)dx  \nonumber \\
&&+\int_{-\infty }^{\infty }g(x,t)\left( \hat{\phi}_{+1}^{\dagger }(x)\hat{%
\phi}_{-1}^{\dagger }(x)e^{-i2\mu t}+{\rm h.c.}\right) dx,  \label{eq2}
\end{eqnarray}
where, $\hat{\phi}_{\pm 1}$ are quantized fields describing the atoms in the
internal levels $|\pm 1\rangle $ satisfying the commutation relation $[\hat{%
\phi}_{i}(x,t),\hat{\phi}_{j}^{\dagger }(x^{\prime },t)]=\delta _{ij}\delta
(x-x^{\prime })$. In Eq.\ (2), the first term describes the kinetic and
potential energy of the atoms in the internal levels $|\pm 1\rangle $. The
second term describes the creation (or annihilation) of two atoms in those
levels due to a collision between two atoms in the condensate. The
condensate is described by a macroscopic wave function and $\mu $ is the
corresponding chemical potential. The function $g(x,t)$ is proportional to
the s--wave scattering length and the condensate density. We have included
the time--dependence explicitly to account for the depletion of the
condensate, as well as to take into account the change of the external
parameters which allow to control the process (\ref{eq1}). On the other
hand, in writing Eq. (2) we have ignored the collisions among the atoms in
levels $|\pm 1\rangle $ as well as the quantum fluctuations of the
condensate, which is valid if the number of atoms in the condensate is large
and in the other levels is small \cite{castdum}. There are other collisional
terms, like $\hat{\phi}_{i}^{\dagger }\hat{\phi}_{i}\phi _{0}^{\ast }\phi
_{0}$ ($i=\pm 1$), which are allowed by the collisional selection rules and are of the
same order in the number of condensate atoms as the second term in the
Hamiltonian. However, they can be included as an effective potential for the
atomic beam fields. In the following, we understand $V\left( x\right) $ as a
renormalized potential which includes such collisional terms.

The quadratic Hamiltonian (2) gives rise to the following linear Heisenberg
equations 
\begin{eqnarray}
i\partial _{t}\hat{\phi}_{\pm 1}(x,t) &=&-\frac{\partial _{xx}^{2}}{2m}\hat{%
\phi}_{\pm 1}(x,t)+V\left( x\right) \hat{\phi}_{\pm 1}(x,t)  \nonumber \\
&&+g\left( x,t\right) \hat{\phi}_{\mp 1}^{\dagger }(x,t)e^{-i2\mu t}.
\label{eq3}
\end{eqnarray}
In the Heisenberg picture, the initial state of the atoms $|\Psi \rangle $ in levels 
$|\pm 1\rangle $ is the vacuum, since we assume that initially (at time $%
t\rightarrow -\infty $) all the atoms are in the condensate, that is, $\hat{%
\phi}_{1,2}(x,-\infty )|\Psi \rangle =0$. Thus, the coupled equations (\ref
{eq3}) describe the generation of atoms in levels $|\pm 1\rangle $ out of
the vacuum. Given a potential $V(x)$ and the function $g(x,t)$ one can solve
them numerically in the same way as one solves the Bogoliubov--de Gennes
equations for the time--dependent excitations of a condensate \cite{castdum}%
. Thus, the solutions can be written in terms of time--dependent Bogoliubov
transformations, which implies that there will be correlations between the
atoms created in levels $|\pm 1\rangle $. This is precisely the physical
origin of the entanglement that will be described in the following.

We will now study two limiting situations: (i) {\em The Squeezing limit}, in
which the typical time for the atoms to leave the trap is much smaller than
the one related to the collisional process, so that many atoms accumulate in
the levels $|\pm 1\rangle $ before they escape from the trap. (ii) {\em The
qubit Entanglement} limit, where the atoms leave the trap before a new pair
of atoms is created. The first situation is analogous to the squeezed light
generation by non-degenerate parametric down conversion \cite{NOPA}, whereas
the second case is similar to the one in which pairs of polarization
entangled photons are created \cite{paradown}.

Let us first study the situation of squeezed atomic beams. We consider a
simple model for which we can obtain the properties of the atomic
correlations analytically. The conclusions drawn from this model are
qualitatively valid for more complicated situations, in which one has to
rely on numerical calculations. We take a potential $V(x)=0$ for $x>0$ and
infinite otherwise, so that $\hat{\phi}_{\pm 1}(0,t)=0$, and assume that the
condensate wave function is such that $g(x,t)=g_{0}(t)$ for $0\leq x\leq a$
and zero otherwise. The function $g_{0}(t)$ is switched on and off in a time
of the order $\gamma ^{-1}$. This time can be related, for example, to the
typical depletion time of the condensate. We will consider separately the
two spatial regions:(I) $x\geq a$; (II) $0<x<a$ and then connect them via
the requirement that the field operators and the first derivatives have to
be continuous.

For $x\geq a$ (I) we can write the solutions of Eq.(\ref{eq3}) as 
\begin{equation}
\hat{\phi}_{\pm 1}(x,t)=\int_{0}^{\infty }\left( \hat{A}_{\pm 1}(\omega
)e^{-ik(\omega )x}+\hat{B}_{\pm 1}(\omega )e^{ik(\omega )x}\right)
e^{-i\omega t}d\omega ,  \label{eq4}
\end{equation}
where $k(\omega )=\sqrt{2m\omega /\hbar }$. Scattering theory assigns the
operators $\hat{A}_{\pm 1}(\omega )$ and $\hat{B}_{\pm 1}(\omega )$ a
definite physical meaning. They are annihilation operators of particles in
incoming and outgoing plane wave modes with velocity $\mp \sqrt{2\hbar
\omega /m}$, respectively. The condition $\hat{\phi}_{\pm 1}(x,-\infty )|\Psi
\rangle =0$ is then translated into $\hat{A}_{\pm 1}(\omega )|\Psi \rangle
=0 $. The physical interpretation is that initially there are no input
atomic beams, so the input modes should be in the vacuum state. The output
modes $\hat{B}_{\pm 1}(\omega )$, determined by the inputs $\hat{A}_{\pm
1}(\omega ) $ and the dynamics in the condensate region, are directly
related to measurable quantities. The state of the output components can be
detected by velocity-selective light imaging \cite{13}. In order to
determine the state of the output modes for vacuum inputs, we need to solve
the dynamical equation (\ref{eq3}) in the condensate region $0\leq x\leq a$.

For this purpose, it is convenient to take a Fourier transformation of Eqs. (%
\ref{eq3}), obtaining a coupled set of equations for $\hat{\Phi}_{\pm
1}(x,\Delta )$ defined through 
\begin{equation}
\hat{\phi}_{\pm 1}(x,t)=e^{-i\mu t}\int_{-\infty }^{\infty }\hat{\Phi}_{\pm
1}(x,\Delta )e^{\mp i\Delta t}d\Delta .  \label{eq5}
\end{equation}
Due to the fact that $g_{0}\left( t\right) $ is time dependent,
the Fourier components $\hat{\Phi}_{+1}(x,\Delta )$ are correlated with 
$\hat{\Phi}_{-1}(x,\Delta -\omega )$ for a range of frequencies $\omega $.
For applications, however, it is desirable to have pure entanglement between 
{\it two} measurable modes, which in our case are the output Fourier
components. In fact, that can be obtained in the limit $\gamma \ll g_{0}$,
where $g_{0}$ is the maximum value reached by $g_{0}(t)$. As shown below,
the bandwidth of $\hat{\Phi}_{\pm 1}(x,\Delta -\omega )$ is roughly
determined by the coupling rate $g_{0}$, which is much larger than the width
of the Fourier transform of $g_{0}(t)$. This means that the Fourier
transform of $g_{0}(t)$ can be replaced by $g_{0}\delta (\omega )$. We call
this approximation the steady output condition, since it corresponds the
physical condition that the atomic loss in the condensate is negligible
before we get a steady output. Imposing the boundary conditions allows us to
express the outgoing operators in terms of the ingoing ones as a Bogoliubov
transformation 
\begin{equation}
\hat{B}_{\pm 1}(\mu \pm \Delta )=\alpha _{\pm 1}(\Delta )\hat{A}_{\pm 1}(\mu
\pm \Delta )+\beta _{\pm 1}(\Delta )\hat{A}_{\mp 1}^{\dagger }(\mu \mp
\Delta ),  \label{eq6}
\end{equation}
where the coefficients $\alpha ,\beta $ can be  determined by solving
the corresponding scattering equations. Note that to keep the commutation
relations these coefficients satisfy the general requirements $|\alpha _{\pm
1}(\Delta |^{2}-|\beta _{\pm 1}(\Delta )|^{2}=1$ and $\alpha _{+1}(\Delta
)\beta _{-1}(\Delta )-\alpha _{-1}(\Delta )\beta _{+1}(\Delta )=0$. From Eq.
(\ref{eq6}), we see that the outgoing modes $\hat{B}_{\pm 1}(\mu \pm \Delta
) $ are in a pure two-mode squeezed state \cite{NOPA}, with the squeezing
parameter $r_{\Delta }$ given by 
\begin{equation}
\tanh \left( r_{\Delta }\right) =\frac{|\beta _{+1}(\Delta )|}{\left| \alpha
_{+1}(\Delta )\right| }=\frac{|\beta _{-1}(\Delta )|}{\left| \alpha
_{-1}(\Delta )\right| }.  \label{eq7}
\end{equation}
The dependence of the squeezing $r_{\Delta }$ on the detuning $\Delta $
determines the squeezing spectrum. Note that a pure two-mode squeezed state
is an ideal continuous variable entangled state, with the entanglement
characterized by the squeezing parameter \cite{9}. Continuous variable
entangled states have many application in recent quantum information
protocols \cite{ce}.

To simplify the expression for the squeezing parameter $r_{\Delta }$, we
assume $\mu \gg g_{0}$. This can be achieved in practice since $\mu $ is
adjustable by changing the shift of level $|0\rangle $. In this case, the
squeezing parameter can be written in the following simple form 
\begin{equation}
r_{\Delta }=\left| 
\mathop{\rm arctanh}%
\left\{ \tanh (2\theta )\sin \left[ \left( k_{+}-k_{-}\right) a\right]
\right\} \right| ,  \label{eq8}
\end{equation}
where $\theta =%
\mathop{\rm arctanh}
[ [ (\Delta /g_{0}) ^{2}+1]^{1/2}-\Delta /g_{0}],$ and $k_{\pm }=[2m\mu
/\hbar \pm (2mg_{0}/\hbar )[(\Delta /g_{0})^{2}+1]^{1/2}]^{1/2}$. The
solution (\ref{eq8}) reveals some interesting properties of this
interaction. First, let us consider a vanishing detuning $\Delta =0$, that is, we look
at the squeezing $r_{0}$ between the output modes $\hat{B}_{+1}(\mu )$ and $%
\hat{B}_{-1}(\mu )$. The parameter $r_{0}$ is given by $r_{0}=\left| 
\mathop{\rm arctanh}%
\left[ \sin \left( g_{0}\overline{t}\right) \right] \right| $, where $%
\overline{t}=2a/\sqrt{2\hbar \mu /m}$ is the transmission time of the input
atomic beam with velocity $\sqrt{2\hbar \mu /m}$ in the region $0\leq x\leq
a $. If the dimensionless interaction coefficient $\kappa =g_{0}\overline{t}%
=\pi /2,$ we have infinite squeezing and infinite output atomic flux. This
simply means that the approximation of negligible atomic loss for the
condensate has broken down at this point. So, similar to the non-degenerate
parametric down conversion in the optical case \cite{NOPA}, our system has a
working threshold given by $\kappa =\pi /2+n\pi $. The system should operate
not very close to the threshold to get steady output of entangled atomic
beams.

Next, let us look at the squeezing spectrum. The squeezing $r_{\Delta }$
versus the dimensionless detuning $\Delta /g_{0}$ and the interaction
coefficient $\kappa =g_{0}\overline{t}$ is shown in Fig.\ 2. From the figure,
we see that we have a broadband two-mode squeezed state with the squeezing
bandwidth roughly determined by $g_{0}$. The steady output condition
requires $g_{0}\gg \gamma $. In our case, the atomic loss is mainly caused
by the output coupling. From Eqs. (\ref{eq4}) and (\ref{eq8}), the loss rate
can be estimated as $\gamma \sim 2g_{0}\sinh ^{2}\left( r_{0}\right) /N_{0}$,
where $N_{0}$ is the total atom number in the condensate. Even with a high
peak squeezing $r_{0}$, the steady output condition can still be easily
attained. It is also interesting to note that one can control the
transmission time $\overline{t}$ by changing the level shift to get a large
peak squeezing $r_{0}$, and we know that the steady output condition does
not put a stringent requirement on the obtainable value of $r_{0}$. Thus, in
this system in principle one can get a much larger squeezing and therefore a
much larger entanglement than in the optical system. As an example, a conservative
estimate gives $g_{0}\sim 20$kHz, $a\sim 3\mu $m, $%
\overline{v}=\sqrt{2\hbar \mu /m}\sim 9$cm/s, corresponding an output flux
about $680$ atoms/ms, we have a very large squeezing $r_{0}\sim 2$, which is
not yet obtainable in current optical systems. The advantage of large
obtainable squeezing in this system is due to the fact that we have a strong
nonlinear interaction caused by the collisions with the condensate. 
\begin{figure}[tbp]
\caption{Squeezing parameter $r$ versus dimensionless detuning $\Delta
/g_{0} $ and interaction coefficient $\protect\kappa =g_{0}\overline{t}$ }
\end{figure}

We emphasize that despite the simple model for the
potential and condensate shape, we expect that all these features will be
present for more realistic models. In fact, if the potential is
asymptotically flat we can write Eq.\ (\ref{eq4}) in that region, so that
under the steady output condition we will obtain Eq.\ (\ref{eq6}) with
different coefficients $\alpha $ and $\beta $. For demonstration of these
features, we suggest a three-step experiment. First, one can demonstrate the
existence of a threshold by controlling the velocity of the output beams.
For a certain velocity of the beams, the spin relaxation rate of the
condensate will increase dramatically, and that gives the threshold value.
Secondly, one can measure the squeezing spectrum by the velocity selective
light imaging \cite{13}. Finally, one can demonstrate the entanglement
between the atomic beams by atomic homodyne detection \cite{9}. This can be
achieved by a atomic beam splitter and a local oscillator provided by an atom laser,
 which can be outcoupled from the same condensate.

Now, we will study the situation in which pairs of atoms are created
sequentially. We will show that, as in the case of photons \cite{paradown,10}%
, if we post-select the measurement results, the corresponding internal
state of the atoms is effectively maximally entangled. For this purpose, we
assume the condensate is located at the region $-a\leq x\leq a$, and the
potential $V\left( x\right) $ is symmetric and independent of the internal
states so that the $|\pm 1\rangle $ atoms have the same probability of going
to the regions $x<-a$ and $x>a$. Two atomic detectors are placed, one on the
left ($x<-a$) and one on the right ($x>a$) of the condensate. The coupling $%
g(x,t)$ in the Hamiltonian (\ref{eq2}) is assumed to be sufficiently small
such that there are at most one pair of atoms generated in each detection
interval. Using perturbation theory (in the Schr\"{o}dinger picture) we can
obtain from the Hamiltonian (\ref{eq2}) the effective atomic state for each
detection interval: 
\begin{equation}
|\Psi (t)\rangle =\int f(x,y,t)\hat{\phi}_{+1}^{\dagger }(x)\hat{\phi}%
_{-1}^{\dagger }(y)dxdy|\text{vac}\rangle ,  \label{eq9}
\end{equation}
where $f(x,y,t)$ can be easily calculated and we have neglected the vacuum
component since it has no influence on the measurement results. After a time 
$t_{0}$,\ the atomic pair leaks out of the condensate, and the wave function 
$f(x,y)\equiv f(x,y,t_{0})\approx 0$ for $-a\leq x,y\leq a$. We can
decompose the wave function in the form $f\left( x,y\right) =f_{LR}\left(
x,y\right) +f_{RL}\left( x,y\right) +f_{LL}\left( x,y\right) +f_{RR}\left(
x,y\right) $, where $f_{LR}\left( x,y\right) $\ is defined to be equal to $%
f\left( x,y\right) $ if $x<-a$ and $y>a$, and to be zero elsewhere. Other
components are defined in a similar way. So the state $\left| \Psi \left(
t_{0}\right) \right\rangle $ is decomposed into four components, with
definite physical meaning for each component. For instance, the component $%
f_{LL}\left( x,y\right) $ represents both of the atoms go to the left side.
Now we project the state onto the subspace where there is one atom at each
side. This projection can be easily achieved in experiments by
post-selections of the measurement results, similar to many optical
experiments involving spontaneous parametric down conversion \cite
{paradown,10}. After the projection, we only have two components in the
effective state $\left| \Psi _{eff}\right\rangle $ (the state selected by
the measurement). The potential is independent of the internal states of the
atoms, so we have $f_{LR}\left( x,y\right) =f_{RL}\left( y,x\right) $. With
this condition, the effective state has the form (not normalized) 
\begin{eqnarray}
\left| \Psi _{eff}\right\rangle &=&\int f_{LR}\left( x,y\right)
\\
&&\left[ \hat{%
\phi}_{+1}^{\dagger }\left( x\right) \hat{\phi}_{-1}^{\dagger }\left(
y\right) +\hat{\phi}_{-1}^{\dagger }\left( x\right) \hat{\phi}_{+1}^{\dagger
}\left( y\right) \right] dxdy |\text{vac}\rangle,   \nonumber
\end{eqnarray}
which can also be written as $\left| +1,-1\right\rangle _{LR}+\left|
-1,+1\right\rangle _{LR}$ with a different notation. Thus with post-selection
of the measurement results, we get an effective maximally entangled state
between the atomic pair, and this state should have many applications in the
field of quantum information, such as measurement of Bell inequalities with
massive particles \cite{aspect}.

In summary, we have analyzed a scheme for generating both entangled atomic
beams and entangled atomic pairs. We have shown that we can get pure
continuous entangled state with a large entanglement and effective maximal
qubit entanglement with post-selection. The generated pure atomic
entanglement can be directly used in the demonstration of many interesting
quantum information protocols \cite{8}.

We thank R. Blatt and J. Anglin for discussions.
A. S.
thanks the university of Innsbruck for hospitality during his visit.
This work was supported by
the Austrian Science Foundation, the Europe Union project EQUIP, the ESF,
the European TMR network Quantum Information, Thomas B. Thriges Center for
Kvanteinformatik, and the Institute for Quantum Information Gmbh.

{\bf Note:} Upon completion of this work, we have become aware of the very
recent preprint quant-ph/0007012 by H. Pu and P. Meystre, which discusses
similar ideas on creating squeezed atomic states.

\end{document}